\documentclass[letterpaper, 10 pt, conference]{ieeeconf}  

\IEEEoverridecommandlockouts                              

\title{\LARGE \bf
Coarse-to-Fine Learning for Multi-Pipette Localisation in Robot-Assisted \textit{In Vivo} Patch-Clamp
}

\author{Lan Wei, Gema Vera Gonzalez, Phatsimo Kgwarae, Alexander Timms, Denis Zahorovsky, \\ Simon Schultz, and Dandan Zhang 
\thanks{*Additional project details are available on our website: \textit{https://lannwei.github.io/Patch-Clamp-Localisation/}}
\thanks{
Lan Wei, Gema Vera Gonzalez, and Dandan Zhang are with the Imperial-X Initiative, Department of Bioengineering, Imperial College London, London, United Kingdom.
Phatsimo Kgwarae, Alexander Timms, Denis Zahorovsky and Simon Schultz are with the Department of Bioengineering, Imperial College London, London, United Kingdom.
Corresponding: Dandan Zhang, {\tt\small d.zhang17@imperial.ac.uk}}
}
\usepackage{amssymb}
\usepackage{xcolor}
\usepackage{comment}

\usepackage[utf8]{inputenc}
\usepackage{url}
\usepackage{booktabs}
\usepackage{bbding}
\usepackage{pifont}
\usepackage{utfsym}
\usepackage{fontawesome}
\usepackage{multirow}
\usepackage{arydshln}
\usepackage{amsmath} 
\usepackage{soul}
\begin{document}
\maketitle
\thispagestyle{empty}
\pagestyle{empty}



\begin{abstract}
\textit{In vivo} image-guided multi-pipette patch-clamp is essential for studying cellular interactions and network dynamics in neuroscience. However, current procedures mainly rely on manual expertise, which limits accessibility and scalability.  Robotic automation presents a promising solution, but achieving precise real-time detection of multiple pipettes remains a challenge. Existing methods focus on \textit{ex vivo} experiments or single pipette use, making them inadequate for \textit{in vivo} multi-pipette scenarios.
To address these challenges, we propose a heatmap-augmented coarse-to-fine learning technique to facilitate multi-pipette real-time localisation for robot-assissted \textit{in vivo} patch-clamp. More specifically, we introduce a Generative Adversarial Network (GAN)-based module to remove background noise and enhance pipette visibility. We then introduce a two-stage Transformer model that starts with predicting the coarse heatmap of the pipette tips, followed by the fine-grained coordination regression module for precise tip localisation. To ensure robust training, we use the Hungarian algorithm for optimal matching between the predicted and actual locations of tips.
Experimental results demonstrate that our method achieved $>98\%$ accuracy within 10\,$\mu m$, and $>89\%$ accuracy within 5\,$\mu m$ for the localisation of multi-pipette tips. The average MSE is 2.52 \,$\mu m$.
\end{abstract}




\section{INTRODUCTION}
The electrophysiology method `whole-cell patch-clamp' is considered the benchmark in neuroscience for recording the electrical activity of cells within neuronal networks~\cite{neher1976single}. By using a glass micropipette to gain intracellular access, this method enables low-noise, high-temporal resolution recordings and precise manipulation of neuronal signals~\cite{suk2019advances,annecchino2018progress}. Simultaneous patch-clamp recordings from multiple neurons increase throughput and enable direct assessment of neuronal connectivity. This approach offers unmatched spatial and temporal precision, providing valuable insights into how neurons interact and function.
The development of \textit{in vivo} patch-clamp enabled scientists to link the behaviour of individual cells to memory formation, perception, decision-making and other cognitive functions in living organisms.
Its unparalleled accuracy makes it indispensable for studying neural function at the cellular level.

To achieve these recordings in intact tissue, significant expertise is required to avoid blood vessels and to compensate for tissue deformation during both navigation and the sealing/break-in of the cell membrane when applied to anaesthetised or awake animals.
The steep learning curve of the technique has hindered its widespread adoption, with only a handful of laboratories worldwide able to perform it.
Furthermore, even among experienced practitioners with extensive training, researchers are typically able to record from only about ten cells per day, with whole-cell success rates varying widely ~\cite{campagnola2022local}. 

\begin{figure}[t!]
\centering
\includegraphics[width=1.0\hsize]{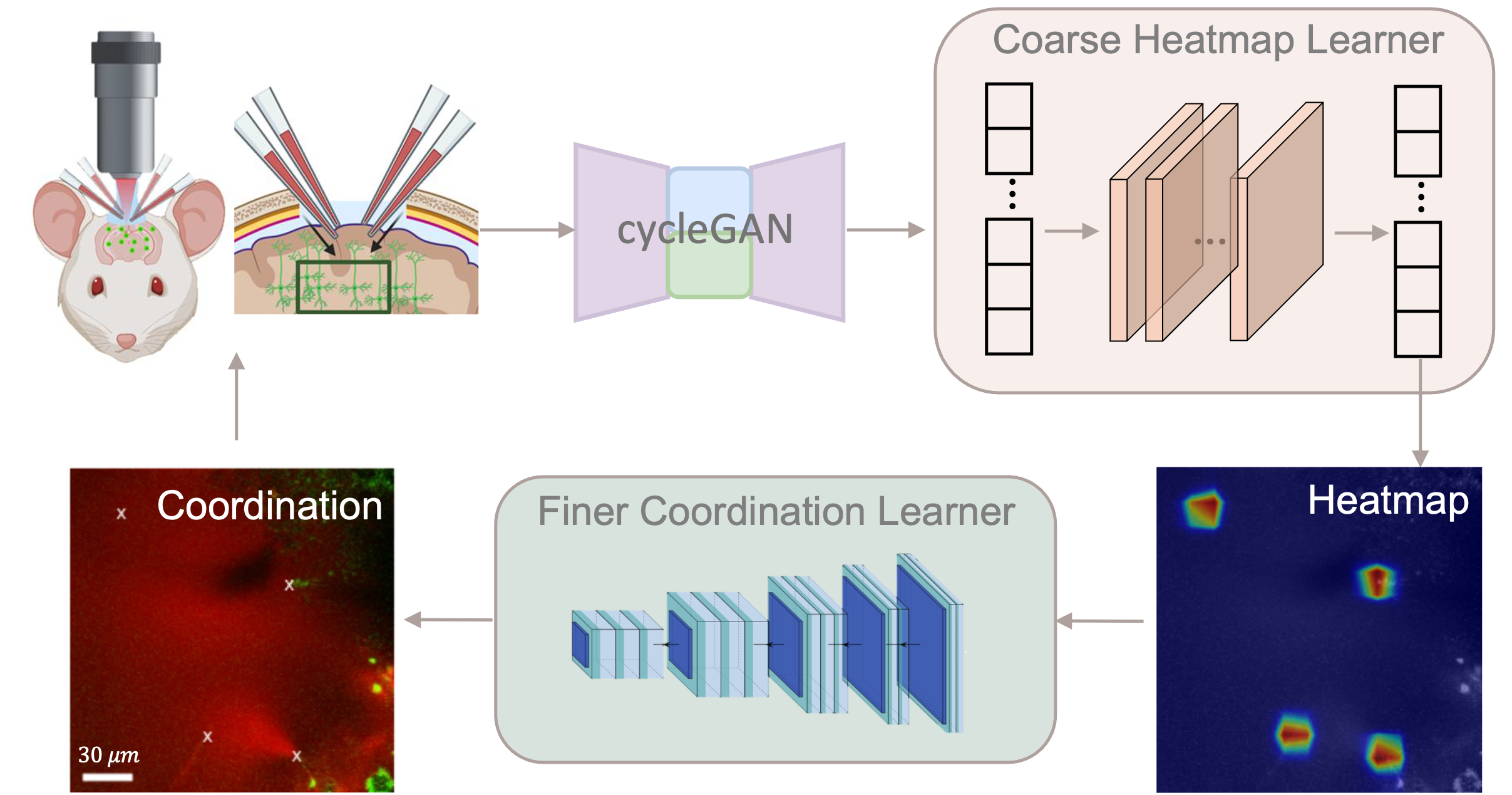}
\vspace{-0.6cm}
\caption[]{Concept overview of our coarse-to-fine learning approach for \textit{in vivo} multi pipettes localisation under two-photon microscopy.
The white cross "$\times$" in the left corner image indicates the position of the pipette tips.}
\label{fig-motivation}
\vspace{-0.6cm}
\end{figure}%

These limitations have driven the development of \textit{in vivo} robotic automation systems, which can be broadly categorised into two architectural paradigms based on targeting methodology:
1) Blind patch-clamp relies on pipette impedance changes to detect and establish contact with random neurons in the vicinity of the pipette tip~\cite{kodandaramaiah2012automated}. 
This approach circumvents the need for real-time imaging but sacrifices cell-type specificity.
2) Image-guided patch-clamp uses optical imaging (most commonly two-photon microscopy) to identify and locate specific neurons, enabling the precise targeting of defined cell types. This is crucial for understanding the function of distinct neuronal populations and for connectivity studies \cite{Margrie2003}.

While robot-assisted targeted methods have significantly increased throughput compared to manual single pipette operations~\cite{suk2017closed,annecchino2017robotic}, their implementation complexity differs drastically between paradigms.
Dual or triple concurrent automated blind whole cell recordings achieved success rates of 29\% in anesthetised and 18\% in awake animals within 10-minute windows~\cite{kodandaramaiah2018multi}. However, no solution for automated image-guided multi-patch-clamp currently exists, partly due to the low resolution and noisy images in two-photon microscopy.

Despite the existing challenges, accurate pipette detection is crucial for successful automated targeted patch-clamp recordings, especially in multi-pipette recording settings.
The technical rationale is twofold:
First, fluorescent dye-filled pipettes require precise initial coordinate establishment to define optimal trajectories toward target cells.
These coordinates allow the micromanipulator’s control system to geometrically estimate pipette tip positions during navigation, especially when they are outside the imaging field. Second, positional verification is mandatory before contact attempts, as repeated tissue penetrations induce cumulative localisation errors in the micromanipulator’s reference frame. Given that the soma of most cells ranges between 10–15\,$\mu m$ in diameter, even minor deviations can result in missing the target entirely.



In brain slice experiments, the issue mentioned above can be mitigated by minimising the number of pipette movements~\cite{gonzalez2023two} or tracking the pipette position~\cite{koos2021automatic,s23198144,gonzalez2021machine}, which is easier in high-contrast \textit{ex vivo}  microscopy images and when there is only one pipette. 
However, \textit{in vivo} applications require frequent adjustments to navigate around blood vessels, making it necessary to iteratively refine the trajectory of the pipette to successfully reach the target cell, which makes multi-patching more challenging~\cite{suk2017closed,annecchino2017robotic}. 
While deep learning-based methods have shown promise for object detection in biological imaging, existing pipette localisation techniques often struggle with low contrast, occlusions, and the need for precise tip positioning. 
Current solutions either lack robustness in noisy environments or require extensive manual intervention, limiting their practical deployment.




To overcome these limitations, we propose a heatmap-augmented coarse-to-fine localisation framework that integrates a Transformer-based encoder with a ResNet-based decoder.
Our hybrid architecture specifically addresses three critical needs: enhanced spatial awareness for two-photon imaging in complex microscopy environments, increased robustness against imaging noise in living tissue, and precise pipette tip coordinate estimation for robot-assisted multi-pipette patch-clamp operations.
The main contributions of this paper are as follows:
\begin{enumerate}
    \item We propose a novel coarse-to-fine localisation framework that employs a Transformer-based encoder to coarsely identify pipette tip regions first, followed by a ResNet-based decoder that precisely predicts their exact coordinates and improves pipette tips detection accuracy.
    \item To mitigate the issues caused by dynamic, low-resolution two-photon images in living organisms, we developed a cycleGAN-based model that effectively eliminates background noise and enhances pipette tip visibility, improving detection accuracy.
    \item We introduce an innovative permutation-invariant loss function, leveraging the Hungarian algorithm to construct a bipartite graph between predicted and ground-truth pipette tip locations, ensuring optimal matching for efficient and robust model training.
    \item We evaluate our method on \textit{in vivo} patch-clamp data with a varying number of pipettes (1 to 4), demonstrating superior performance over existing approaches.

\end{enumerate}


\section{RELATED WORK}

\vspace{-0.1cm}
\subsection{Ex Vivo patch-clamp Localisation}
Koos et al.~\cite{koos2021automatic} developed an \textit{ex vivo} pipette tip detection method based on geometry analysis of dark regions obscured by the pipette. 
Li et al.~\cite{s23198144} employed a Mixture of Gaussian algorithm to enhance single pipette detection by focusing on the tip region under the microscope. 
Gonzalez et al.~\cite{gonzalez2021machine} introduced a ResNet-based deep learning model for single pipette tip localisation in brain slice images, demonstrating the potential of deep learning for automated patch-clamp recordings. While useful, \textit{ex vivo} studies are limited compared to \textit{in vivo} because they lack the physiological context of intact brain systems, including blood flow and inter-regional connectivity, and are prone to changes in tissue properties over time. These factors make \textit{ex vivo} findings less representative of real conditions~\cite{opitz2017limitations}.

\begin{table}[!t]
\caption{Comparison of automatic patch-clamp pipette localisation.}
\vspace{-0.2cm}
\centering
\begin{tabular}{c|c|c|c}
\hline\hline
Method & \textit{In Vivo} & Multi-Pipette & Real Time \\ \hline 
\cite{koos2021automatic} &   \ding{55}  &  \ding{55}  &     Unknown   \\ 
\cite{gonzalez2021machine}  & \ding{55} & \ding{55}  &  \checkmark    \\ 
\cite{s23198144} &  \ding{55}  &  \ding{55}  &  \checkmark  \\ 
\cite{long20153d}& \checkmark & \ding{55} &  Unknown \\ 
\cite{suk2017closed}& \checkmark & \ding{55} & \checkmark \\  \hline
 \textbf{Ours} & \checkmark  &  \checkmark  & \checkmark  \\ \hline\hline
\end{tabular}
\label{table:relate}
\vspace{-0.5cm}
\end{table}

\begin{figure*}[t!]
\centering
\includegraphics[width=1.0\hsize]{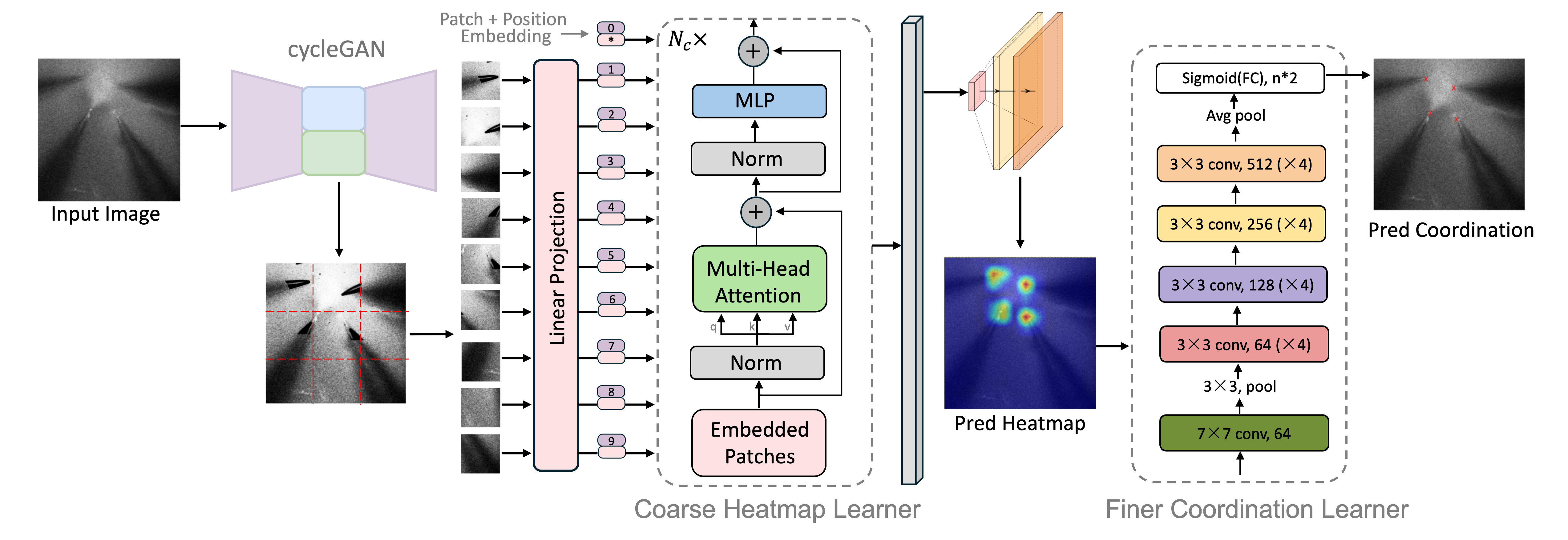}
\vspace{-0.5cm}
\caption[]{Framework Overview. The model takes the original \textit{in vivo} image as input and first applies cycleGAN to enhance pipette tip features. The enhanced image, augmented with positional embeddings, is then processed by the encoder, which predicts a coarse heatmap of pipette tip locations. Finally, the decoder refines these predictions, generating precise pipette tip coordinates through the finer coordination learner.
}
\label{fig-framework}
\vspace{-0.5cm}
\end{figure*}%

\vspace{-0.1cm}
\subsection{In Vivo patch-clamp Localisation}

Two-photon laser scanning microscopy was integrated into \textit{in vivo} patch-clamp setups to enable the imaging of fluorescent signals from deep brain regions.
The first automated system for \textit{in vivo} patch-clamp recordings, developed by Kodandaramaiah et al.~\cite{kodandaramaiah2012automated}, was designed for blind patching, meaning it lacked visual guidance for targeting specific cells.
Building on this advancement, Long et al.~\cite{long20153d} pioneered the first attempt to automate image-guided patching \textit{in vivo}, enabling a single patch pipette to be automatically localised near a targeted neuron.
However, the initial localisation of the pipette tip still required manual intervention by the experimenter.
Subsequently, Suk et al.~\cite{suk2017closed} introduced a real-time pipette localisation method that automated single-pipette tracking by analysing consecutive image frames to identify clusters of fluorescently labelled bright pixels corresponding to the pipette tip.
This innovation improved targeting precision and reduced the need for manual intervention in single-pipette patch-clamp recordings.


Nevertheless, existing single-pipette detection methods are difficult to apply directly to multi-pipette scenarios in real-time applications, 
due to increased spatial complexity and the need for simultaneous tracking of multiple moving pipettes.
Therefore, we propose a multi-stage neural network that processes two-photon microscopy images with the following innovations:
1)	Feature Enhancement – Enhancing pipette tip visibility to improve detection accuracy;
2)	Coarse-to-Fine localisation – A hierarchical prediction approach that refines tip coordinates for precise positioning.
As shown in Table~\ref{table:relate}, our method is the first to achieve real-time multi-pipette localisation \textit{in vivo}, marking a significant advancement in automated patch-clamp technology.

\section{METHODOLOGY}

The workflow of the proposed method for multi-pipette tips localisation is illustrated as follows (as shown in Fig.~\ref{fig-framework}).
\begin{enumerate}
    \item Step 1:
To reduce the domain gap between the patch-clamp image of \textit{ex vivo} data and \textit{in vivo} data, the CycleGAN~\cite{zhu2017unpaired} model is used to translate the pipette images obtained from the live brain (\textit{in vivo}) with a noisy background to the target domain (\textit{ex vivo}) with a clear background.
    \item Step 2: A Vision Transformer (ViT) (encoder model) is used to predict the heatmap of the pipette tips' coarse positions.
    \item Step 3: The generated heatmap is fed to a ResNet-based decoder model to obtain the tips' coordination.
\end{enumerate}
\subsection{Noise Background Elimination via GAN}
\begin{figure*}[t!]
\centering
\includegraphics[width=1.0\hsize]{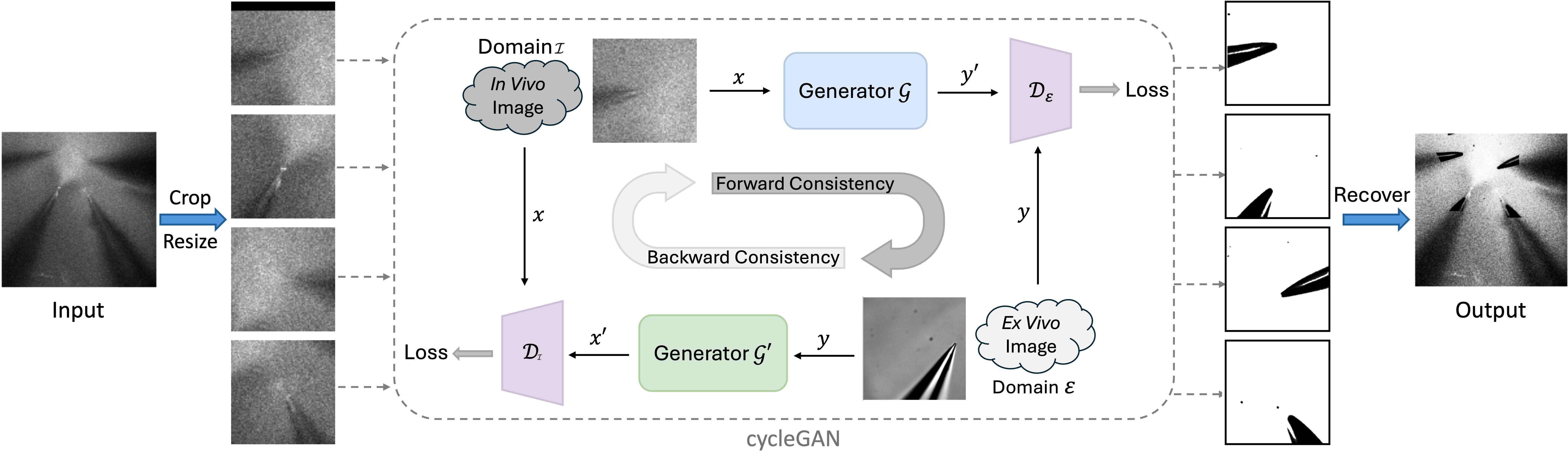}
\vspace{-0.5cm}
\caption[]{Workflow of cycleGAN-based pipette tip feature enhancement. 
A single pipette is first cropped from the \textit{in vivo} two-photon microscopy images. 
CycleGAN is then applied to remove background noise, enhancing pipette tip visibility. 
Finally, the denoised single pipette images are mapped back to their original positions while preserving the original pipette tip coordinate labels.
}
\label{fig-gan}
\vspace{-0.5cm}
\end{figure*}%
Let $\mathcal{E}$ represent \textit{ex vivo} patch-clamp data from brain slices that have a clear background and an explicit pipette boundary, and $\mathcal{I}$ represent \textit{in vivo} patch-clamp data from live tissue, which is of lower resolution. 
To eliminate background noise in $\mathcal{I}$, we employ a CycleGAN, which consists of a generator and a discriminator trained in an adversarial manner, as shown in Fig.~\ref{fig-gan}. 
Specifically, the generator \(\mathcal{G}\colon \mathcal{I} \rightarrow \mathcal{E}\) aims to transform \textit{in vivo} images to resemble those in the \textit{ex vivo} domain, while the discriminator \(\mathcal{D}_{\mathcal{E}}\) learns to distinguish the generated images \((\mathcal{E}^{\prime})\) from real \textit{ex vivo} images \((\mathcal{E})\). 
Conversely, the inverse generator \(\mathcal{G}^{\prime}\colon \mathcal{E} \rightarrow \mathcal{I}\) maps \textit{ex vivo} images back to the \textit{in vivo} domain, and its discriminator \(\mathcal{D}_{\mathcal{I}}\) ensures that the translated images \((\mathcal{I}^{\prime})\) are indistinguishable from real \textit{in vivo} images.

To ensure that transformations preserve essential information, we impose a cycle consistency loss $\mathcal{L}_{\text{cyc}}(\mathcal{G}, \mathcal{G}{\prime})$. 
This loss compels the model to accurately reconstruct the images for each domain when cycled through both generators $(\mathcal{I} \approx \mathcal{G}^{\prime}(\mathcal{G}(\mathcal{I}))$ and $\mathcal{E} \approx \mathcal{G}(\mathcal{G}^{\prime}(\mathcal{E})))$.
The optimisation is formulated as a min-max problem:
\begin{equation}
    \mathcal{G}^*, \mathcal{G}^{\prime *} = \arg\min_{\mathcal{G},\mathcal{G}'} \max_{\mathcal{D}_\mathcal{I},\mathcal{D}_\mathcal{E}} \mathcal{L}(\mathcal{G}, \mathcal{G}', \mathcal{D}_\mathcal{I}, \mathcal{D}_\mathcal{E})
\end{equation}


Given that each \textit{in vivo} image contains multiple pipettes and each \textit{ex vivo} image contains only one pipette, we first isolate individual pipettes from \textit{in vivo} images by cropping. 
This is achieved by extracting a fixed-size (80 $\times$ 80) patch centered around each labelled pipette tip, applying padding when necessary to maintain uniform dimensions.
We then apply the trained generator $\mathcal{G}$ to each cropped image, thereby removing background noise. 
Finally, the denoised single-pipette images are mapped back to their original positions, carrying over the original pipette tip coordinate labels, which are used for the following pipettes' localisation.

\subsection{Pipettes Localisation}

Following the attention mechanism, the Transformer processes each image as a sequence of patches and captures global dependencies for representation. 
While a global view is beneficial for vision tasks, research has shown that local context modeling is also essential for dense-level vision tasks~\cite{li2021localvit}. 
For our localisation task, particularly in scenarios involving multiple pipettes with ambiguous boundaries, global dependency aids in identifying the coarse positions of pipette tips but lacks the necessary local context to precisely determine their exact coordinates.
Therefore, we propose to find the tip information in a coarse-to-fine Transformer model to explore the local context of tips' coordination.

Given an input image \( I \in \mathbb{R}^{H \times W \times 3} \) with width \( W \) and height \( H \), it is first divided into non-overlapping patches of size \( 16 \times 16 \). These patches are then flattened into 1D patch embeddings and augmented with a learnable positional embedding~\cite{gehring2017convolutional}, which is randomly initialized to compensate for the spatial information lost during sequentialisation. This process results in the final sequential embedding \( f^0 \in \mathbb{R}^{\frac{H}{16} \times \frac{W}{16} \times d} \), where \( d \) is the embedding dimension.

Localisation is achieved using a sequence of $N_{c}$ layers in a coarse area learner (\(\text{co}\)) followed by $N_{f}$ layers in a finer coordination learner (\(\text{fi}\)) to refine the extracted features. The process is formulated as:
\vspace{-0.1cm}
\begin{equation}
    \hat{f}, \xi \longleftarrow \mathcal{F}_{\text{fi}}^{1 \ldots N_{f}}\left(\mathcal{F}_{\text{co}}^{1 \ldots N_{c}}\left(f^0\right)\right),
\end{equation}
where the predicted heatmap \( \hat{f} \in \mathbb{R}^{H \times W} \) represents the coarse pipette tip locations, and \( \xi \in \mathbb{R}^{n \times 2} \) denotes the precise coordinates of the \( n \) detected tips.

\subsubsection{Attention-based Coarse Heatmap Learner}
The coarse learner aims to constrain the model's attention on the surrounding area of the pipette tips as they contribute more to the final tips localisation result.
With this inspiration, we propose to use the self-attention module to find such areas in the manner of predicting the tips' heatmap.
Specifically, it contains $N_{c} = 12$ Vision Transformer~\cite{dosovitskiy2020image} encoder blocks in total.
Assumed that at the $i-$th block, given the input feature as $f^{i-1}$, we first feed it into a sequence of multi-head self-attention (MSA) and multilayer perception (MLP) to gather the global dependency for coarsely locating the tip of the pipettes. 
After each part, there is a Layer Normalisation with residual short connection for a stable training process~\cite{vaswani2017attention}.
We denote this intermediate feature as:
\vspace{-0.2cm}
\begin{equation}
    z^i=\mathcal{F}_{M S A}\left(f^{i-1}\right) \oplus \mathcal{F}_{M L P}\left(\mathcal{F}_{M S A}\left(f^{i-1}\right)\right),
\end{equation}
where $\oplus$ denotes element-wise addition. 
We replace the classification head of the original ViT with an identity module to preserve the 768-dimensional feature map. 
Then, the heatmap head is used to generate the tips heatmap $\hat{f}$ which consists of a 3×3 convolution that reduces the feature dimensionality to 256, followed by a ReLU activation. Next, a bilinear upsampling operation increases the spatial resolution by a factor of 16, restoring it to the input scale. Finally, a 1×1 convolution layer outputs a single-channel map, thus producing the high-resolution heatmap that highlights the tips of interest.


\subsubsection{Attention-based Finer Coordination Learner}

We employ a ResNet18 backbone for finer coordination regression by modifying its first layer to accept a single-channel heatmap as input. Additionally, we replace the final classification head with a fully connected layer that outputs the coordinates of the \( n \) detected tips.
During the forward pass, the network processes the heatmap through the ResNet backbone and generates raw coordinate predictions. 
We then apply a sigmoid activation function and scale the output by 256 to map the normalised coordinates back to the input image size. 
Finally, the predictions are reshaped to obtain the \((x, y)\) coordinates \( \xi \) of each detected tip.

\subsection{Heatmap Generator}
\label{heatmap_gen}

We introduce an algorithm to generate the ground-truth key-patch heatmap for training the encoder with full supervision.
The heatmap is generated by drawing a Gaussian distribution centered at the key tip points.
The equation used to generate the Gaussian distribution for each point is as follows:
\vspace{-0.2cm}
\begin{equation}
    \operatorname{Gaussian}(x, y)=\exp \left(-\frac{\left(x-x_0\right)^2+\left(y-y_0\right)^2}{2 \sigma^2}\right),
\end{equation}
where $x_0, y_0$ are the centre coordinates and $\sigma$ is the standard deviation controlling spread.
The heatmap of the image $I$ with $n$ pipette is generated by:
\vspace{-0.2cm}
\begin{equation}
    \operatorname{Heatmap}_I= \sum_{i=1}^n\operatorname{Gaussian}(x^i, y^i).
\end{equation}

\subsection{Matches between True and Prediction}
To find a match between predicted key points and their corresponding ground truth locations, we begin by computing a cost matrix ($c \in \mathbb{R}^{n\times n}$) where each row represents a predicted key point and each column represents a ground truth key point. 
The entries of this matrix reflect the cost (e.g., Euclidean distance) of assigning a given predicted point to a specific true point. 
We then apply the Hungarian (Munkres) algorithm~\cite{bougleux2017hungarian} to this cost matrix, which provides an optimal one-to-one assignment ($a \in \mathbb{R}^{n\times 2}$) between the predicted and ground truth points by minimising the total overall cost:
\begin{equation}
    \min \sum_{i=1}^n \sum_{j=1}^n c_{i j}a_{i j}.
\end{equation}
This assignment not only handles differences in the number of predictions versus ground truth points but also ensures that each predicted point is paired to the most appropriate ground truth point before calculating the subsequent loss.

\subsection{Objective Function}
To train the localisation network including the proposed two learner blocks, we employ two types of functions.
The first one is the Dice loss function $\phi_{DICE}(\cdot)$ to minimise the difference between the ground truth heatmap and the predicted heatmap as $\mathcal{L}_{\text{m}}$.
The second one is a designed Hungarian distance loss $\phi_{Hungarian}$ to reduce the predicted tips coordination and its ground truth as $\mathcal{L}_{\text{h}}$.
Total loss is defined as:
\vspace{-0.2cm}
\begin{equation}
\begin{gathered}
\mathcal{L}=\mathcal{L}_{\text{m}}+ \alpha \sum_{i=1}^{n} \mathcal{L}_{\text{h}}^i, \label{loss_fun}\\
\mathcal{L}_{\text{m}}=\phi_{DICE}\left(f_{gt}, \hat{f}_{\text{pred}}\right),\\
\mathcal{L}_{h}^i=\phi_{Hungarian}\left(\xi_{gt}^i, \hat{\xi}_{\text {pred}}^i\right),
\end{gathered}
\end{equation}
where $\alpha \in (0, 1)$, $\hat{\xi}_{\text {pred}}^i$ denotes the $i-$th predicted tip coordination and $n$ denotes the total number of pipettes which is set to 4 in this paper.

\section{EXPERIMENTS}
\begin{figure}[!t]
\centering
\includegraphics[width=1\hsize]{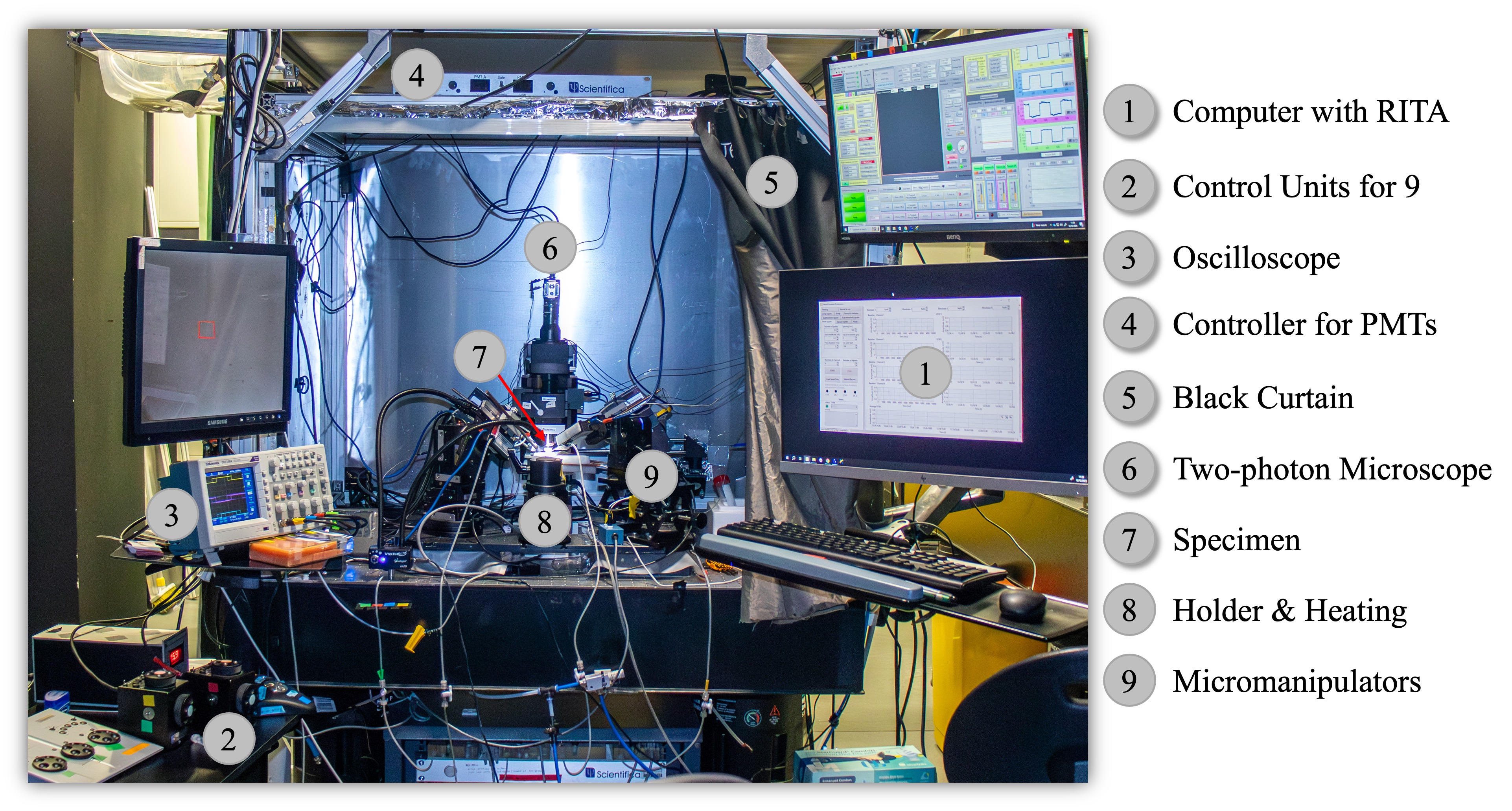}
\vspace{-0.4cm}
\caption[]{Physical setup of our robot-assisted \textit{in vivo} two-photon targeted multi-patch-clamp system.
RITA~\cite{gonzalez2023two} is the software that controls the robot and
PMTs are the microscopy receptors of light.
}
\label{fig-invivo}
\vspace{-0.55cm}
\end{figure}%

\subsection{\text{In vivo} experimental setup}
Using the system shown in Fig.~\ref{fig-invivo}, we conducted \textit{in vivo} patch-clamp experiments on adult male and female C57BL/6J wild-type mice (4–12 weeks old). 
For targeted patching, fluorescent protein expression was induced via AAV PHP.eB CAG-NLS-GFP (Addgene catalogue 104061) injection into the facial vein at postnatal day zero. 
The surgical procedure for patch-clamp recording involved a scalp incision, skull exposure, and removal of connective tissue, followed by headplate implantation using cyanoacrylate glue and dental cement to ensure stability and electrical isolation. 
A craniotomy (2–3 mm) was performed above the right cortex, the dura was carefully dissected, and saline was applied to keep the brain surface moist. 
Anaesthesia was induced and maintained throughout the surgical and experimental sessions using an injectable fentanyl-midazolam-medetomidine cocktail (2~µg/mL fentanyl, 0.2~mg/mL midazolam, 20~µg/mL medetomidine). The induction dose was 0.5~ml per 20~g body weight, with lower volume top-ups administered every 1-2 hours as needed.
The animals were placed on a heating pad to maintain a body temperature of approximately $37^\circ\text{C}$, with their heads mechanically fixed to a custom-built stereotaxic frame for recording under the microscope.

Patch pipettes, with a resistance of 5–7.5~$M\Omega$, were pulled beforehand from borosilicate glass capillaries, loaded into holders, and checked for cleanliness and correct resistance before being positioned above the brain surface. 
Pipette visualisation was achieved using a fluorescent dye (Alexa Fluor 594) excited under 810–880~nm two-photon (2P) excitation.  
The recording platform consisted of a Newport/Spectra-Physics MaiTai HP laser, a multi-photon microscope (Scientifica Ltd.), and a micromanipulator per channel (Sensapex or Scientifica), each fitted with mechanical stability clamps to minimise vibration, tissue deformation, and pressure fluctuations.

\subsection{Datasets}
\begin{figure}[!t]
\centering
\includegraphics[width=0.75\hsize]{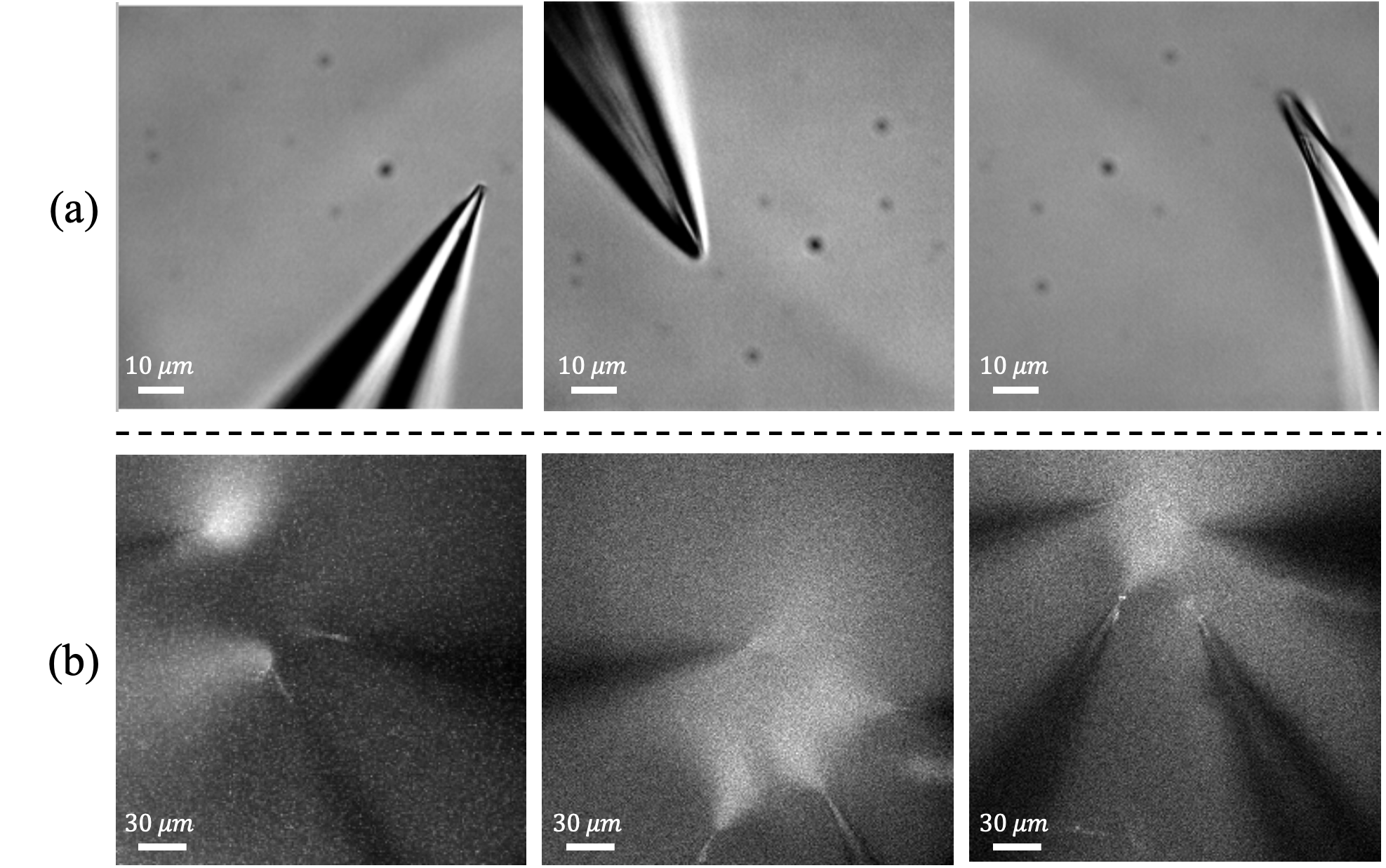}
\caption[]{Visualisation of two datasets. (a) 3 examples from \textit{ex vivo} Dataset~\cite{gonzalez2021machine} featuring a single pipette with a clear background. (b) 3 examples from our in-house \textit{in vivo} dataset, captured using two-photon microscopy, containing multiple pipettes.}
\label{fig-dataset}
\vspace{-0.55cm}
\end{figure}%

For the proposed algorithm, we used two datasets, examples of which are shown in Fig.~\ref{fig-dataset}.
The first dataset consists of 6,678 images of a single pipette with a resistance of 3–5~$M\Omega$ (tip diameter: 1–2 mm)~\cite{gonzalez2021machine}, captured against a plain background and within a brain slice
using a standard electrophysiology setup (SliceScope Pro 3000, Scientifica Ltd) with PatchStar micromanipulators at a 24$^\circ$ approach angle.
This dataset serves as the target domain for training the CycleGAN model.


The second in-house \textit{in vivo} dataset was captured using a two-photon microscopy quad-pipette setup~\cite{gonzalez2023two} (as shown in Fig.~\ref{fig-invivo}) and consists of 341 multi-pipette images.
Each image contains 1–4 pipettes with a resistance of 5–7.5~$M\Omega$, and the position of each pipette tip is labelled by an expert at the start of a multi-patch-clamp trial.
This dataset serves as the source domain for training the CycleGAN model.

\begin{figure*}[!t]
\centering
\includegraphics[width=1.0\hsize]{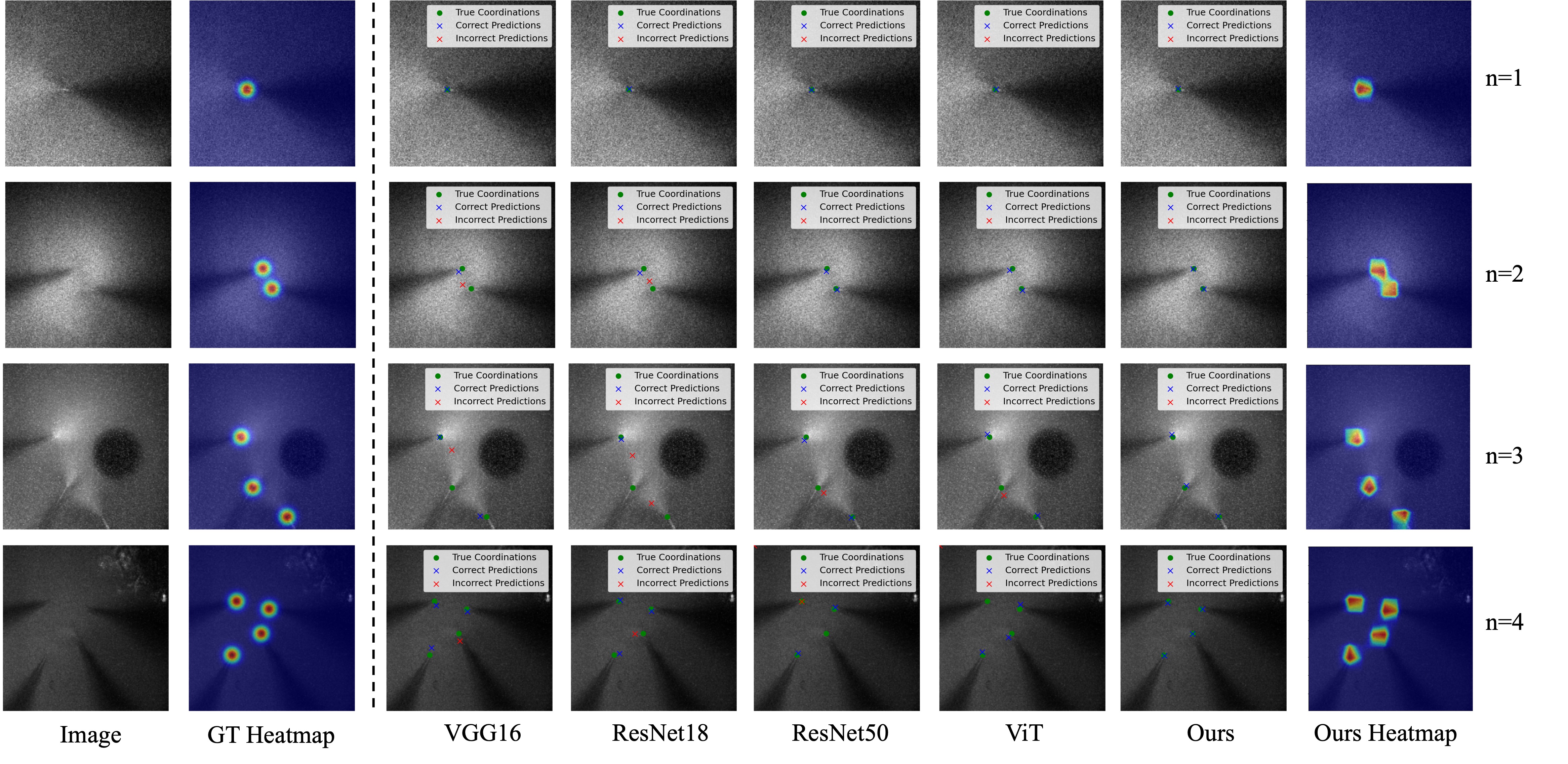}
\vspace{-0.7cm}
\caption[]{Comparison of pipette localisation across images with varying numbers of pipettes (1 to 4). Green points indicate the ground-truth pipette tip positions, while blue crosses (\textcolor{blue}{$\times$}) represent correctly predicted tips within an error range of $10\,\mu m$ and red crosses (\textcolor{red}{$\times$})  denote incorrect predictions.}
\label{fig-sota}
\vspace{-0.3cm}
\end{figure*}%

\subsection{Implementation Details}
All images are resized to \( 256 \times 256 \), and data augmentation is applied using vertical flips, horizontal flips, and random scaling within the range of 0.9–1.1. 
Training is performed with a batch size of 32 using the Adam optimiser.
After cropping to get the single pipette image \( 80 \times 80 \) and data augmentation, we use 6,584 images to train cycleGAN for 200 epochs with an initial learning rate of 0.0002 for the first 100 epochs and linearly decay learning rate to zero in the following 100 epochs. The pipette localisation encoder (ViT) and decoder (ResNet-18) has been pre-trained on ImageNet. 


We employ a multi-stage model training strategy. First, we freeze the decoder and train only the heatmap prediction encoder model for 50 epochs with a learning rate of 0.0001. 
Next, we freeze the encoder and train the coordinate prediction decoder model for 50 epochs with a learning rate of 0.001.  
Finally, we unfreeze both encoder and decoder and train them together for 100 epochs at a learning rate of 0.0001. We reduce the learning rate by half whenever the validation loss does not improve for 20 consecutive epochs. 
The parameter $\alpha$ in equation (\ref{loss_fun}) is set to 0.15, and the  standard deviation $\sigma$ used for heatmap generation is set to 10, 
determined through hyperparameter tuning over 20 rounds using the Optuna tool~\cite{akiba2019optuna}. For the baseline models, we fine-tune the pre-trained models available in \textit{torchvision}~\cite{torchvision2016} for 200 epochs. 


\subsection{Evaluation Metrics}

We evaluate the accuracy of pipette localisation using three key metrics. 
First, we assess localisation accuracy quantitatively by computing the Euclidean distance between predicted and ground-truth pipette tip coordinates under minimal matching. 
For qualitative evaluation, we calculate the percentage of correctly predicted points, where the Euclidean distance to the ground truth falls below a predefined error threshold \( e \). 
In our experiments, we set \( e \) to 3~$\mu m$, 5~$\mu m$, and 10~$\mu m$.
We evaluate the accuracy of heatmap-based localisation using the Intersection over Union (IoU) metric, measuring the overlap between the predicted heatmap and the ground-truth heatmap.


\subsection{Comparison with State-of-the-Art}

\begin{table}[!t]
\centering
\caption{Comparison with State-of-the-Arts.
}
\vspace{-0.3cm}
\begin{tabular}{c|cccc}
\hline \hline
\multirow{2}{*}{Model} &\multirow{2}{*}{MSE$\downarrow$} &e = 3 $\mu m$ & e = 5 $\mu m$&e = 10 $\mu m$\\
\cdashline{3-5}
 &   & Acc.\%$\uparrow$  & Acc.\%$\uparrow$ & Acc.\%$\uparrow$\\ \hline 
VGG16~\cite{simonyan2014very} & \underline{11.70}& 31.72 & 49.65&69.65\\
ResNet18~\cite{he2016deep}& 13.83 &17.24 &42.75& 68.96\\
ResNet50~\cite{he2016deep} &18.22 & \underline{55.86} &\underline{73.79}&81.37\\
ViT~\cite{dosovitskiy2020image} & 19.34&35.86 &62.06& \underline{84.13}\\ \hline
Ours &\textbf{2.52} & \textbf{73.79} &\textbf{89.65} & \textbf{98.62}\\
 \hline \hline
\end{tabular}
Note: The best results in each column are highlighted in \textbf{bold} and the second‐best are \underline{underlined}.
$\downarrow$ indicates that the lower the better and $\uparrow$ the opposite. 
\label{table:sota}
 \vspace{-0.3cm}
\end{table}

To compare performance, we conduct experiments using both convolutional-based models (VGG16~\cite{simonyan2014very}, ResNet18~\cite{he2016deep}, ResNet50~\cite{he2016deep}) and a transformer-based model (ViT~\cite{dosovitskiy2020image}).
Table~\ref{table:sota} highlights the significant improvements achieved by our model across all baselines in terms of quantitative metrics. 
Even at the smallest error threshold of \( 3~\mu m \), our method outperforms the best baseline by 17.93\%, demonstrating its superior prediction accuracy.

Fig.~\ref{fig-sota} provides a visual comparison of pipette tip localisation across images containing varying numbers of pipettes (1 to 4).
The first row represents the simplest scenario, featuring a single pipette, where all models accurately predict the tip position.
However, as the number of pipettes increases, baseline models struggle to correctly identify multiple tip locations, whereas our model maintains high localisation accuracy.
The last column of Fig.~\ref{fig-sota} illustrates the heatmap predictions generated by our model, further highlighting its precision in detecting pipette tips.

\subsection{Inference Time Comparison}
As shown in Table~\ref{table:time}, we evaluated the forward inference time of our model across different hardware configurations (CPU, GPU, and TPU) by running 100 iterations on a batch of images. 
The purpose of this comparison is to demonstrate the applicability of our model to different experimental setups with varying levels of computational power, highlighting the real-time performance that is crucial for robot-assisted \textit{in vivo} patch-clamp.
To ensure accurate measurement, we first warm up the model, synchronise CPU/GPU operations, and then compute the average elapsed time per iteration, normalised by batch size. 
Our results indicate that the multi-pipette tip localization process takes an average of 0.07 seconds on the CPU and 0.001 seconds on the RTX 4090 GPU. Resutls demonstrate the real-time capability of our method across various computing environments.

\begin{table}[!t]
\centering
\caption{Inference time across different hardware devices. 
}
\vspace{-0.2cm}
\begin{tabular}{c|c:ccc:c}
\hline \hline
\multirow{2}{*}{Hardware} & \multirow{2}{*}{CPU} & Titan X & RTX 4090  & Colab T4 & Colab v2-8 \\ 
 & & GPU & GPU  & GPU & TPU \\ \hline
Time (s) & 0.0764 & 0.0064 & 0.0013  & 0.0109 & 0.0011\\ \hline \hline
\end{tabular}
Note: The second column reports results using a CPU-only setup, while the next three columns correspond to different GPU types. 
The final column presents inference time using a TPU available on Google Colab.
\label{table:time}
\vspace{-0.5cm}
\end{table}

\subsection{Ablation Study}
To evaluate the contribution of each component in our model, we conduct ablation studies, as presented in Table~\ref{table:ablation}, under an error threshold of 10\,$\mu m$. Specifically, we examine the effectiveness of the following three key components:
1) GAN-based feature enhancement for improving pipette visibility; 
2) Coarse learner for heatmap-based tip localisation 
and 3) Finer learner for coordinate regression and precise tip positioning.
The results demonstrate that even GAN-based feature enhancement alone already outperforms the baseline models (comparison of the first row in Table~\ref{table:ablation} with the best baseline in Table~\ref{table:sota}), highlighting its effectiveness in improving localisation accuracy.
Among all components, the finer learner (third column) contributes the most significant improvement, as it accurately refines the pipette tip location from the extracted heatmap.
In summary, Table~\ref{table:ablation} confirms that each of our design choices positively impacts the overall performance, with the combination of all components achieving the highest accuracy.
\begin{table}[!th]
\centering
\caption{Ablation of different blocks under error range of 10\,$\mu m$.}
\begin{tabular}{ccc|ccc}
\hline \hline
\multirow{2}{*}{GAN} & Coarse  & Finer & \multirow{2}{*}{IoU$\uparrow$}  & \multirow{2}{*}{MSE$\downarrow$} & \multirow{2}{*}{Acc.\%$\uparrow$}  \\ 
&Learner&Learner&&& \\ \hline 
\checkmark &\ding{55} &\ding{55}&- & 15.08& 84.82 (+0.69)\\
\checkmark & \checkmark  & \ding{55} &0.62 &5.24 & 89.74 (+4.92)\\
\checkmark &\ding{55} & \checkmark&- &4.67 & 93.58 (+3.84)\\
\checkmark & \checkmark  & \checkmark&\textbf{0.69} &\textbf{2.52} &\textbf{98.62} (+5.04) \\
 \hline \hline
\end{tabular}

Note: Since the 1-st and 3-rd lines refer to the model without a coarser learner, they do not predict the heatmap and the IoU value is nan. The numbers in parentheses (+XX) in the last column represent the improvement in accuracy compared to the model that has one less attribute.
\label{table:ablation}
\vspace{-0.5cm}
\end{table}

\section{Discussion and Future Work}
\subsection{Model Design Discussion}

For pipette tip feature enhancement via GAN, we utilize cycleGAN~\cite{zhu2017unpaired} due to its ability to perform unsupervised image translation without requiring paired training data. 
Although Pix2Pix GAN~\cite{isola2017image} has been developed for pixel-level image translation, it relies on paired training data, which is infeasible for our noise background elimination task since images from \textit{ex vivo} and \textit{in vivo} domains cannot be directly paired. 
CycleGAN, on the other hand, introduces a cycle-consistency loss, enforcing an inverse mapping from the target domain back to the source domain, ensuring that translated images retain essential structural information. This makes cycleGAN well-suited for reducing the domain gap between \textit{in vivo }and \textit{ex vivo} data in multi-pipette patch-clamp localisation, allowing for effective background noise suppression while preserving pipette tip features.

For the finer coordination regression module backbone selection, we evaluated four models: a Transformer-based decoder~\cite{fujitake2024dtrocr}, a CNN model, a YOLO-based object detector~\cite{tian2025yolov12}, and a ResNet-based pre-trained model~\cite{he2016deep}, with ResNet performing best. 
The Transformer-based decoder struggles to capture fine-grained local features, which are crucial for precise pipette tip localization. Moreover, it requires a larger training dataset to generalise well in low-resolution two-photon images. 
The three-layer CNN model, which takes the heatmap as input, lacks deep feature representation, making it less robust in complex and noisy backgrounds. 
The YOLO-based model, originally designed for natural object detection, is not well-suited for heatmap-based coordinate regression. Moreover, even the smallest YOLOv12 model (6.5 GFLOPS) is significantly more computationally expensive than our ResNet-based model (1.74 GFLOPS), which limits its suitability for real-time inference.

In contrast, the ResNet-based model, with its skip connections, effectively preserves both low-level and high-level features, ensuring robust pipette tip detection. 
Furthermore, leveraging a pre-trained ImageNet model enhances generalization, particularly in cases where labeled training data is limited. 
The combination of residual learning and pre-trained feature extraction provides superior performance by balancing local detail preservation with global structure recognition, resulting in higher accuracy in multi-pipette coordinate regression.

\subsection{Future Work}
In future work, we aim to integrate multi-modal signal processing into our image-based localisation algorithm to further enhance robot-assisted automatic patch-clamp systems. 
In addition to visual information, we will incorporate electrophysiological signals, such as changes in voltage ($V$) and resistance ($M\Omega$), which indicate contact between the pipette and the neuron.
Currently, all two-photon \textit{in vivo} images used in our approach capture pipettes before they enter the brain, similar to certain \textit{ex vivo} pipette localisation systems, where localisation is more straightforward. 
However, to fully automate the patch-clamp process, we plan to develop our algorithm to operate inside brain tissue, where challenges such as tissue deformation and optical distortions arise. 
Additionally, we aim to extend our localisation framework beyond 2D (X-Y) to include depth (Z), enabling full 3D pipette tracking for more precise and robust targeting.

\vspace{-0.03cm}
\section{Conclusions}
In this work, to the best of our knowledge, we introduce the first real-time multi-pipette localisation algorithm for robot-assisted \textit{in vivo} patch-clamp, addressing the challenges posed by low-resolution, high-noise two-photon microscopy images.
By integrating GAN-based feature enhancement with a coarse-to-fine localisation framework, our method first employs a Transformer-based encoder to coarsely identify pipette tip regions, followed by a ResNet-based decoder that precisely predicts their exact coordinates.
Our approach significantly outperforms baseline methods, achieving $>98\%$ accuracy within $10\,\mu m$ and $>89\%$ accuracy within $5\,\mu m$, while operating in real time.
This advancement enhances the scalability and precision of automated \textit{in vivo} multi-patch-clamp, facilitating more efficient electrophysiology studies across diverse neural recording applications.


\section*{Acknowledgement}
Lan Wei acknowledges support from a PhD studentship jointly sponsored by the Imperial-CNRS PhD Programme, the China Scholarship Council, and the Department of Bioengineering, Imperial College London.
Gema Vera Gonzalez acknowledges support from the Eric and Wendy Schmidt AI in Science Postdoctoral Fellowship, a Schmidt Futures program.
Simon Schultz acknowledges support from EPSRC EP/W024020/1. 

All experimental procedures were carried out in accordance with the Animals (Scientific Procedures) Act 1986 and followed Home Office and institutional guidelines under a Project License issued to S.R. Schultz (PP6988384). 
We acknowledge that Biorender was used to create part of the figure.
\vspace{-0.1cm}


\bibliographystyle{IEEEtran}
\bibliography{main}
\end{document}